The LaTex file follows:

\documentstyle[art12,bezier]{article}

\begin{document}

\renewcommand{\thefootnote}{\fnsymbol{footnote}}

\thispagestyle{empty}

\hfill \parbox{45mm}{
{MPI-PhT/94-53} \par
August 1994}

\vspace*{15mm}

\begin{center}
{\LARGE Potential Energy in Quantum Gravity.}

\vspace{22mm}

{\large Giovanni Modanese}%
\footnote{A. Von Humboldt Fellow.}

\medskip

{\em Max-Planck-Institut f\"ur Physik \par
Werner-Heisenberg-Institut \par
F\"ohringer Ring 6, D 80805 M\"unchen (Germany)}

\bigskip

\end{center}

\vspace*{20mm}

\renewcommand{\thefootnote}{\arabic{footnote}}
\setcounter{footnote} 0
\begin{abstract}
We give a general expression for the static potential energy of the
gravitational interaction of two massive particles, in terms of an invariant
vacuum expectation value of the quantized gravitational field. This formula
holds for functional integral formulations of euclidean quantum gravity,
regularized to avoid conformal instability. It could be regarded as the
analogue of the Wilson loop of gauge theories and allows in principle, through
numerical lattice simulations or other approximation techniques,
non perturbative evaluations of the potential or of the effective
coupling constant.

\bigskip \bigskip

\end{abstract}

\newcommand{\beq}{\begin{equation}}    \newcommand{\la}{\langle}
\newcommand{\eeq}{\end{equation}}      \newcommand{\ra}{\rangle}
\newcommand{\beqa}{\begin{eqnarray}}   \newcommand{\pa}{\partial}
\newcommand{\eeqa}{\end{eqnarray}}     \newcommand{\half}{\frac{1}{2}}

\newcommand{\m}{\medskip}

\section{Introduction.}

The present paper is concerned with the problem of the energy of the
gravitational field. This energy has been under investigation
since the birth of General Relativity and some issues, like the
determination of the total energy of a field configuration, have been
settled in a rigorous way in the ADM formalism \cite{dewitt} or
through Noether's theorem \cite{jacki2}. Other points, however, like the
possibility of ``localizing'' the gravitational energy, are still
obscure (see \cite{energy} and references therein).

Here we shall concentrate on a special case, namely that of the static
potential energy of the gravitational interaction of two massive bodies.
The gauge-invariant description of this interaction in terms of
quantum fields is a crucial issue for any physically reasonable theory
of quantum gravity.

This potential energy will turn out to be related to the vacuum
average of a simple gauge-invariant functional of the field, a kind of
correlation between ``scalar'' Wilson lines. While evaluation of
this average for a weak field on a flat background yields the Newton
potential in the usual form \cite{vacuum}, non-perturbative evaluations
could show modifications in the effective coupling constant or in
the dependence of the potential on the distance between the particles.

Like in the case of the Wilson loop, our formula can be implemented
quite naturally on a lattice version of the theory, in order to allow
numerical simulations (Hamber and Williams, \cite{hw}). The first results
are very interesting,
showing that in the strong-gravity region of the phase space of lattice
theory the potential is indeed Yukawa-like, with a mass parameter which
decreases towards the critical point (where the average curvature
vanishes). The effective value of Newton constant has been estimated too.

This paper extends and refines an earlier work \cite{energy}. The outline
is the following. In Section 2 we recall in brief the general-relativistic
view about the gravitational potential energy. In Section 3 we give a formula
which, treating two masses as external sources
for a quantized gravitational field on a flat
background, allows to write the potential energy of their interaction. This is
done using a known technique of euclidean quantum field theory \cite{symanz}.
In Section 4 we generalize the formula to the case of ``strong'' gravity,
introducing a definition of the source which avoids the use of
background metric. Finally, in Section 5 we
discuss the compatibility of the background formula with the general
one and the possible behaviour of the effective coupling constant.

\section{Potential Energy Versus ADM Energy.}

In this Section we work in (3+1) dimensions and follow the conventions of
Weinberg \cite{weinbe}. Starting from next Section, the metric will
be euclidean. The Einstein action is given by
\beq
  S_{\rm Einst.} = - \frac{1}{16 \pi G}
  \int d^4x \, \sqrt{g(x)} \, R(x)
\label{skv}
\eeq
and the action of a material particle of mass $m$ is
\beq
  S_{\rm Mat.} = - m \int dp \, \sqrt{-g_{\mu \nu}[x(p)]
  \dot{x}^\mu(p) \dot{x}^\nu(p)} ,
\eeq
where $x^\mu(p)$ is the trajectory of the particle and $p$ is any
parameter. The dots will always denote differentiation with respect to
the parameter. Finally, the Einstein equations have the form
\beq
  R_{\mu \nu} - \half g_{\mu \nu} R = - 8 \pi G T_{\mu \nu} .
\eeq

Let us now decompose the metric in the traditional way
\beq
  g_{\mu \nu}(x) = \eta_{\mu \nu} + h_{\mu \nu}(x)
\eeq
and denote the linearized Einstein equations in the harmonic gauge as
\beq
  K^{\rho \sigma}_{\mu \nu} \, h_{\rho \sigma} = T_{\mu \nu} .
\eeq
The inverse of the kinetic operator $K$ is the well-known
Feynman-De Witt propagator
\beq
  K^{-1}_{\mu \nu \rho \sigma}(x-y) = - \frac{2G}{\pi} \,
  \frac{\eta_{\mu \rho} \eta_{\nu \sigma} +
  \eta_{\mu \sigma} \eta_{\nu \rho} -
  \eta_{\mu \nu} \eta_{\rho \sigma}}{(x-y)^2 - i\epsilon} .
\eeq

\m
Let us compute the Newton potential starting from the preceding equations.
This is an elementary calculation, but the integral one encounters will be
useful in the following.
The field produced by a generic four-momentum source $T_{\rho \sigma}$ in the
linearized approximation is given by
\beq
  h_{\mu \nu}(x) = \int d^4y \, [K^{-1}]_{\mu \nu}^{\rho \sigma}(x-y)
  \, T_{\rho \sigma}(y) ;
\eeq
when the source is a particle of mass $m$ at rest in the origin, the
only non vanishing component of $T_{\rho \sigma}$ is
\beq
  T_{00}(y) = m \, \delta^3({\bf y}) ,
\eeq
so we have
\beqa
  h_{00}(x^0,{\bf x}) & = &
  \int d^4y \, [K^{-1}]^{00}_{00}(x-y) \, m \, \delta^3({\bf y}) \nonumber \\
  & = & m \int_{-\infty}^{+\infty} dy^0 \int d^3y \,
  \frac{\frac{-2G}{\pi} \, \delta^3({\bf y})}
  {({\bf x}-{\bf y})^2-(x^0-y^0)^2-i\epsilon}
  \nonumber \\
  & = & m \int_{-\infty}^{+\infty} dy^0 \,
  \frac{\frac{-2G}{\pi}}{{\bf x}^2-(x^0-y^0)^2-i\epsilon}
  = \frac{2mG}{|{\bf x}|} .
\label{fre}
\eeqa
This is the correct result, since in the newtonian approximation
we have
\beq
  g_{00} = -1 - 2V ,
\eeq
where $V$ is the Newton potential.

More generally, one can use the ADM formula.
In classical General Relativity the total energy (mechanical + gravitational)
of a physical system is given by the ADM mass formula
\beq
  E = - \frac{1}{16 \pi G} \int
  \left( \frac{\pa h_{jj}(x)}{\pa x^i} - \frac{\pa h_{ij}(x)}{\pa x^j}
  \right) n^i r^2 d\Omega ,
\label{sle}
\eeq
where the integral is computed on a surface at spacelike infinity.
We shall briefly review here the connection
between this energy and the static gravitational potential.

A generic static metric $g_{\mu \nu}$ can be written at spatial
infinity in the form
\begin{eqnarray}
  g_{00} & \simeq & -1 + \frac{2M_1 G}{|{\bf x}|} +
  O\left(\frac{1}{|{\bf x}|^2}\right) ; \label{due} \\
  g_{0i} & \simeq & O\left(\frac{1}{|{\bf x}|^2}\right) ; \label{duf} \\
  g_{ij} & \simeq & \delta_{ij} + \frac{2M_2 G}{|{\bf x}|^3} x_i x_j
  + O\left(\frac{1}{|{\bf x}|^2}\right) . \label{dug}
\end{eqnarray}

Performing the integral (\ref{sle}) with $h_{ij}$ given by (\ref{dug})
one sees that $M_2$ is the ADM energy (``total mass''). On the other hand,
$M_1$ is the mass observed by measuring the newtonian force at infinity.
Substituting (\ref{due}) - (\ref{dug}) into Einstein's equations
$R_{\mu \nu}=0$, it is easy to see that $M_1=M_2$. This is a quite
natural result \cite{dewitt}. In other words it means that, according to
special relativity conceptions, the source of the newtonian
field is not only the mass, but also the energy density.
For instance, in the gravitational collapse of a star a part of the
gravitational energy is converted into kinetic energy and eventually
this energy is employed to produce heavier elements from hydrogen or
helium. If we disregard the radiation emitted into space, the newtonian
field far away from the star remains unchanged during the whole process
and the same holds for the ADM mass, which is a conserved quantity.

The gravitational potential energy can be found, by definition, assuming
a static distribution of matter and computing the metric it generates at
infinity. This has been done by Murchada and York \cite{muryor} for a
spherical matter distribution of uniform density, using conformal
transformations and a special formulation of the initial-value equations
of General Relativity. For a sphere of (small) density $\rho$ and unit
radius they found the right newtonian gravitational binding energy,
namely the ADM mass is given in this case by (reintroducing the radius
$R$ and the velocity of light $c$)
\begin{equation}
  M_{\rm TOT} = \frac{4}{3} \pi R^3 \rho - \frac{1}{c^2} \frac{16}{15}
  \pi^2 \rho^2 G R^5 + O(\rho^3) .
\label{xsw}
\end{equation}
Remembering that $M_{\rm TOT}$ also represents the effective source of the
newtonian field, we see that the second term in (\ref{xsw}) gives
rise to a deviation from the famous law which states the independence
of the potential on the radius of the source. Nevertheless, this
effect is usually unobservable, due to the very small factor $c^{-2}$.

It is also possible to find the following corrections to (\ref{xsw}),
proportional to $\rho^3$, $\rho^4$, ... They denote the existence of
general-relativistic corrections to the potential energy $m_1 m_2 G/r$.
For instance, the term proportional to $\rho^3$ would contribute to
$M_{\rm TOT}$ a term of the form
\begin{equation}
  \Delta M \propto \frac{1}{c^4} \rho^3 G^2 R^7 + O(\rho^4).
\label{kgf}
\end{equation}

In the case of a source constituted by two pointlike bodies of masses
$m_1$ and $m_2$, kept at rest at a fixed distance $r$, the method of
solution mentioned above is not applicable.  From eq.\ (\ref{xsw}) we may
infer that the ADM mass is given in this case by
\begin{equation}
  M_{\rm TOT} = m_1 + m_2 - \frac{1}{c^2} \frac{G m_1 m_2}{r} +
  o\left(\frac{1}{c^2}\right).
\end{equation}
There are, however, no relativistic corrections to the
two-body potential coming from (\ref{kgf}), because the corresponding potential
does not admit a continuum limit. For instance, if we try to integrate
a potential of the form $G^2 m_1^{3/2} m_2^{3/2}/r^2$ to obtain the
term proportional to $\rho^3$, we find that the binding energy of the
sphere depends on the way it has actually been put together.
This means that the correction proportional to $\rho^3$ comes from a
three particles potential, and so on for $\rho^4$ etc.

\section{Quantum formula for the potential energy on
a flat background.}

The same result we found in the preceding Section using the classical equations
of motion can be obtained in a completely different way. It is known
that the ground state energy of a system described by an action
$S_0[\phi]=\int d^4x \, L(\phi(x))$ in the presence of external sources $J(x)$
can in euclidean quantum field theory be expressed as
\beqa
  E & = & \lim_{T \to \infty} - \frac{\hbar}{T} \log
  \frac{\int d[\phi] \exp \left\{ - \hbar^{-1} [\int d^4x \, L(\phi(x)) +
  \int d^4x \, \phi(x) J(x) ]\right\}}{\int d[\phi]
  \exp \left\{ - \hbar^{-1} \int d^4x \, L(\phi(x)) \right\} }
  \nonumber \\
  & & \nonumber \\
  & = & \lim_{T \to \infty} - \frac{\hbar}{T} \log
  \left< \exp \left\{ - \hbar^{-1} \int d^4x \, \phi(x) J(x) \right\} \right> ,
\eeqa
where, outside the interval $(-\half T,\ \half T)$, the source has been
switched off.

\m
This equation has been proved exactly in perturbation theory \cite{symanz} for
the case of a linear local coupling between the field and the external source,
but it can be generalized assuming that in any case the vacuum-to-vacuum
transition amplitude is given by
\beqa
  \la 0^+ | 0^- \ra_J & = &
  \frac{\int d[\phi] \exp \left\{ - \hbar^{-1}
  ( S_0[\phi] + S_{\rm Inter.}[\phi,J] ) \right\}}
  {\int d[\phi] \exp \left\{- \hbar^{-1} S_0[\phi] \right\} }
  \nonumber \\
  & & \nonumber \\
  & = &  \left< \exp
  \left\{\hbar^{-1} S_{\rm Inter.}[\phi,J] \right\} \right> .
\eeqa
In fact, inserting a complete set of energy eigenstates we can write
\beq
  \la 0^+ | 0^- \ra_J = \la 0 | e^{-HT/\hbar} | 0 \ra =
  \sum_n \la 0 | e^{-HT/\hbar} | n \ra \la n | 0 \ra =
  \sum_n |\la 0 | n \ra|^2 \, e^{-E_n T/\hbar} .
\eeq
The smallest energy eigenvalue $E_n$ corresponds to the ground state, and in
the limit $T \to \infty$ it dominates the sum. So taking the logarithm and
multiplying by $(-\hbar/T)$ we obtain that energy.
This is a well-known technique in QCD
(see for instance \cite{bander}).

\m
In the case of a weak gravitational field quantized on a flat background, we
may consider the source constituted by two masses $m_1,\ m_2$, placed at rest
near the origin at a distance $L$ each from the other (see eq.\ (\ref{kty})).
The action of this system is
\beqa
  S & = & - \frac{1}{16\pi G} \int d^4x \sqrt{g(x)} \, R(x)
  - m_1 \int_{-\frac{T}{2}}^{\frac{T}{2}} dt_1 \,
  \sqrt{g_{\mu \nu}[x(t_1)] \dot{x}^\mu(t_1) \dot{x}^\nu(t_1) }
  \nonumber \\
  & & - m_2 \int_{-\frac{T}{2}}^{\frac{T}{2}} dt_2 \,
  \sqrt{g_{\mu \nu}[y(t_2)] \dot{y}^\mu(t_2) \dot{y}^\nu(t_2) } ,
\eeqa
where the trajectories $x^\mu(t_1)$ and $y^\mu(t_2)$ of the
particles with respect to the background are simply given by
\beq
  x^\mu(t_1) = \left( t_1,\, -\frac{L}{2},\, 0,\, 0 \right); \qquad
  y^\mu(t_2) = \left( t_2,\, \frac{L}{2},\, 0,\, 0 \right) .
\label{kty}
\eeq

So we have
\beqa
  & & E_{\rm B-G} = \lim_{T \to \infty} - \frac{\hbar}{T} \times \nonumber \\
  & & \ \ \log \frac{\int d[h] \, \exp \left\{ - \hbar^{-1} \left[
  S_{\rm Einst.} +
  m_1 \int_{-\frac{T}{2}}^{\frac{T}{2}} dt_1 \,
  \sqrt{1 + h_{00}[x(t_1)] } +
  m_2 \int_{-\frac{T}{2}}^{\frac{T}{2}} dt_2 \,
  \sqrt{1 + h_{00}[y(t_2)]} \right] \right\} }
  {\int d[h] \, \exp \left\{ - \hbar^{-1} S_{\rm Einst.} \right\} }
  \nonumber \\
  & & \nonumber \\
  & & = \lim_{T \to \infty} - \frac{\hbar}{T} \log
  \left< \exp \left\{ - \hbar^{-1} \left[
  m_1 \int_{-\frac{T}{2}}^{\frac{T}{2}} dt_1 \,
  \sqrt{1 + h_{00}[x(t_1)] } +
  m_2 \int_{-\frac{T}{2}}^{\frac{T}{2}} dt_2 \,
  \sqrt{1 + h_{00}[y(t_2)]} \right] \right\} \right> .
  \nonumber \\ & &
\label{skt}
\eeqa

We call this expression ``B-G'' (background-geometry) formula
\cite{vacuum}. It has the property that the time $T$ is referred
to the background geometry. It follows in particular that the
average is sensitive also to fields $h$ which are pure gauge modes
and carry no curvature. For further details on this point compare
next Section and the discussion in Section 5.

By standard perturbation techniques it is
straightforward to see that for weak fields and to lowest order in $G$
eq.\ (\ref{skt}) reduces to
\beqa
  E_{\rm B-G} & = & m_1+m_2+ \lim_{T \to \infty} - \frac{\hbar}{T} \log
  \left\{ 1 + \frac{m_1 m_2}{4\hbar^2} \int_{-\frac{T}{2}}^{\frac{T}{2}} dt_1
  \int_{-\frac{T}{2}}^{\frac{T}{2}} dt_2 \, \left<
  h_{00}[x(t_1)]
  h_{00}[y(t_2)] \right> \right\}
  \nonumber \\
  & \simeq & m_1+m_2+ \lim_{T \to \infty} - \frac{1}{T} \frac{m_1 m_2}{4}
  \int_{-\frac{T}{2}}^{\frac{T}{2}} dt_1
  \int_{-\frac{T}{2}}^{\frac{T}{2}} dt_2 \,
  K^{-1}_{0000}(\tau_1-\tau_2,\, L,\, 0,\, 0) \nonumber \\
  & = & m_1+m_2 -  \frac{m_1 m_2 G}{L} .
\eeqa
In the last step we have used the classical equation (\ref{fre}).
The next term in the perturbative series would be the first quantum correction,
proportional to $\hbar G^2$.

Eq.\ (\ref{skt}), like the corresponding ones in QED or QCD, has the physically
appealing feature of showing how the force between the sources ultimately
arises from the exchange of massless gravitons. However, let us make a closer
comparison with electrodynamics. In that case the analogue of the functional
integral which appears in the logarithm of (\ref{skt}) has the form
\cite{fischl}
\beq
  \left< \exp \left\{ g \int_{-\frac{T}{2}}^{\frac{T}{2}} dt_1
  A_0[x(t_1)] - g \int_{-\frac{T}{2}}^{\frac{T}{2}} dt_2
  A_0[y(t_2)] \right\} \right> .
\label{vge}
\eeq
(The two charges have been chosen to be opposite: $q_1=g$, $q_2=-g$.) Reversing
the direction of integration in the second integral and closing the contour at
infinity, one is able to show that the quantity (\ref{vge}) coincides with the
Wilson loop of a single charge $g$, thus giving a gauge invariant expression
for the potential energy.

In gravity this is not possible: we may imagine that an expression like
(\ref{vge}) could be obtained in the first-order formalism (with $A_0$ replaced
by the tetrad $e^0_0$), but the masses necessarily have the same sign,
so the loop cannot be closed.

\section{General case.}

We aim now at giving a quantum formula for the potential which does not
rely on a fixed background. We assume that a functional integral
for euclidean gravity exists, denoted by
\beq
  Z = \int d[g] \exp \left\{ - \hbar^{-1} S_{\rm Einst.}[g] \right\}
\label{chh}
\eeq
and require that all the field configurations in this functional
integral are asymptotically flat.

In practice it is known that the euclidean Einstein action is not bounded from
below, so one has to be careful in treating it. The action needs to be
regularized in some way; in the actual lattice simulations, this happens
thank to the $R^2$ term and to the measure (see \cite{hw} and references).

Since we have no background, the quantities $T$ and $L$ which enter in
the definition of the source must now refer to the dynamical metric
$g$, configuration by configuration, in the functional integral (\ref{chh}).

A first, simple recipe for generalizing the B-G equation (\ref{skt})
along this lines is the following. Let us consider two massive
bodies (with $m_1=m_2$, for simplicity) which are kept at a fixed
invariant distance $L$ from each other by some device and ``fall''
through a given field configuration $g$. Suppose that the motion is started
and stopped in the asymptotically flat region. If $T$ is the total
proper time measured along the trajectory of the first mass (with $T \gg L$),
the proper time measured along the trajectory of the other mass will
be, say, $T+\alpha[g]$. Then the action of this system leads,
after integration over all field configurations, to the energy
\beqa
  E_{\rm T-G} & = & \lim_{T \to \infty} - \frac{\hbar}{T} \log
  \left< \exp \left\{ - \hbar^{-1} m (2T + \alpha[g]) \right\} \right>
  \nonumber \\
  & = & 2m + \lim_{T \to \infty} - \frac{\hbar}{T} \log
  \left< \exp \left\{ - \hbar^{-1} m \alpha[g] \right\} \right> .
\label{cde}
\eeqa
We call this expression ``T-G'' (total geometry) formula. Unlike the B-G
formula (eq.\ (\ref{skt})), it refers only to geometrical,
invariant quantities.

If we suppose that the field configurations contributing to the average
are not too singular, then $\alpha$ is small and we may
expand the exponential, finding (note that $\la \alpha[g] \ra$ obviously
vanishes by symmetry)
\beq
  E_{\rm T-G} = 2m + \lim_{T \to \infty} - \frac{m^2}{2\hbar T}
  \la \alpha^2 [g] \ra + ...
\label{poi}
\eeq
Thus the interaction energy is always negative in this case. Also,
eq.\ (\ref{poi}) gives an intuitive picture of the gravitational
attraction: it arises because keeping two masses at some fixed
distance increases their action, due to the vacuum fluctuations
of the gravitational field.

\m
Now, we turn to a more careful description of the whole procedure above,
and eliminate the (apparent! -- thank to the average) asymmetry
between the two bodies, assigning the total proper time $T$ to the
geodesic of their center of mass.

Let us suppose that a field configuration is given. We consider a geodesic
line of length $T$, which starts at an arbitrary point in the ``past''
asymptotically flat region with unit timelike velocity.

To fix the ideas, this curve could be written in its first part as
\beq
  \xi^\mu(\tau) = \left( -\half T + \tau, \, 0,\, 0,\, 0 \right) ;
  \qquad 0 \leq \tau \leq \bar{\tau} ,
\eeq
where $\tau$ is the proper time measured along the curve and we have chosen the
remaining coordinates of the starting point to be equal to $(0,\, 0,\, 0)$
(this
is an irrelevant arbitrariness, since at the end we shall integrate over all
the configurations of the field). As usual, $T$ denotes a very long time
interval. After a time $\simeq \bar{\tau}$ the curve enters the region of
spacetime where the gravitational field is non vanishing. It continues as a
geodesic, which means that $\xi^\mu(\tau)$ satisfies the equation
\beq
  \Gamma^\rho_{\mu \nu}[\xi(\tau)] \dot{\xi}^\mu(\tau) \dot{\xi}^\nu(\tau)
  + \ddot{\xi}^\rho(\tau) = 0,
\eeq
where $\Gamma^\rho_{\mu \nu}$ is the Christoffel symbol of the metric. The
curve terminates at $\tau=\half T$, again in the flat region.

\m
Let us then take in the initial point $\xi^\mu(0)$ a unit vector $q^\mu(0)$,
orthogonal to $\dot{\xi}^\mu(0)$ (for instance, in our example,
$q^\mu(0)=(0,\, 1,\, 0,\, 0)$), and define a vector $q^\mu(\tau)$ along the
curve $\xi^\mu(\tau)$ by parallel transport of $q^\mu(0)$. We remind that
$\dot{\xi}^\mu(\tau)$, being the tangent vector of a geodesic, is parallel
transported along the geodesic itself, and that the parallel transport
preserves the norms and the scalar products. Then the following relations
hold along the curve
\beqa
  \dot{\xi}^\mu(\tau) \dot{\xi}^\nu(\tau) g_{\mu \nu}[\xi(\tau)] & = & -1 ;\\
  q^\mu(\tau) q^\nu(\tau) g_{\mu \nu}[\xi(\tau)] & = & 1 ;\\
  \dot{\xi}^\mu(\tau) q^\nu(\tau) g_{\mu \nu}[\xi(\tau)] & = & 0 .
\eeqa

Next we consider two masses $m_1, \ m_2$, and a length $L$ which we may regard
as infinitesimal, compared to the scale $T$. We assume that the two masses
follow the trajectories $x^\mu(\tau)$ and $y^\mu(\tau)$, respectively, given by
\beqa
  x^\mu(\tau) & = & \xi^\mu(\tau) - L_1 q^\mu(\tau) ; \label{zsw} \\
  y^\mu(\tau) & = & \xi^\mu(\tau) + L_2 q^\mu(\tau) , \label{zss}
\eeqa
where $L_1$ and $L_2$ are two positive lengths such that
\beq
  L_1+L_2 = L \qquad {\rm and} \qquad -m_1 L_1 + m_2 L_2 = 0 .
\label{huy}
\eeq

\m
The physical meaning of the preceding geometrical construction is apparent: it
represents an observer which falls freely in the center of mass of the system
composed by $m_1$ and $m_2$, while holding the two masses at rest at a distance
$L$ each from the other. This is the generalization of the source introduced in
eq.\ (\ref{kty}) that is naturally dictated by the equivalence principle.

We notice that if the two masses were allowed to fall freely in the field, they
would not keep at a constant distance from each other. In fact, as it is well
known from the so-called geodesic deviation equation, the distance between two
neighboring geodesics varies according to the sign of the curvature in the
region they are traversing.

\m
We can reparameterize the two curves $x^\mu(\tau)$ and $y^\mu(\tau)$
introducing their proper times $\tau_1$ and $\tau_2$, respectively.
The ratio between the proper time $\tau_1$ and the proper time $\tau$
is given by the equation
\beq
  d\tau_1 = \sqrt{-g_{\mu \nu}[x(\tau_1)]
  \dot{x}^\mu(\tau_1) \dot{x}^\nu(\tau_1)} d\tau .
\eeq
An analogous relation holds for $\tau_2$. We agree to adjust the function
$\tau_1(\tau)$ in such a way that $\tau_1(0)=0$. Then we shall denote
$\tau_1(-\half T)=-\half T_1'$, $\tau_1(\half T)=\half T_1''$ and
$T_1 = \half (T_1'+T_1'')$. For flat
geometries we have $T_1'=T_1''=T_1=T$. Analogous relations hold for $\tau_2$.

The energy can then be written as
\beq
  E_{\rm T-G} = \lim_{T \to \infty} - \frac{\hbar}{T} \log
  \left< \exp \left\{ - \hbar^{-1} [m_1 T_1 + m_2 T_2] \right\} \right> .
\label{kuy}
\eeq

In order to make contact with eq.\ (\ref{cde}), suppose now
to take $m_1=m_2$ and to ``attach''
smoothly the ends of the curves $x(\tau)$ and $y(\tau)$ to the
geodesic $\xi(\tau)$. The corresponding change of length will have no effect on
eq.\ (\ref{kuy}), because it is in no way a function of $T$. Since
a geodesic has minimal length with respect to neighboring curves,
we have, for the total times $T_1$ and $T_2$ as functions
of $L$
\beqa
  \frac{T_1}{T} & = & 1 + \half a[g] L^2 + \frac{1}{6} b[g] L^3 + ... \\
  & & \nonumber \\
  \frac{T_2}{T} & = & 1 + \half a[g] L^2 - \frac{1}{6} b[g] L^3 + ...
\eeqa
where $a[g]>0$, while $b$ has no definite sign. From this we can find
the difference between $T_1$ and $T_2$, which is the analogue of the
quantity $\alpha[g]$ of eq.\ (\ref{cde}):
\beq
  \frac{T_1}{T_2} = 1 + \frac{1}{3} b[g] L^3 + ... \qquad
  {\rm that \ is} \qquad \frac{\alpha[g]}{T} \sim b[g] L^3 .
\eeq

\section{Discussion.}

The B-G formula (\ref{skt}) and the more general T-G formula (\ref{cde})
(which is the one actually implemented in the lattice simulations)
must be consistent in the limit of small $G$. In this Section we are
going to clarify this point and work out some consequences.

We first recall that expanding the euclidean functional integral (\ref{chh})
from flat space in perturbation theory, one finds that to order
$\hbar G$ all field configurations are pure gauge modes carrying no
curvature.
\footnote{This also suggests that gravitons are, physically, quite
fictitiuos objects.}
In fact, the Wilson loops of the connection vanish to this
order, and this cannot be an average effect, as they have the form
$W=(\theta_1^2+\theta_2^2)$, where $\theta_1$ and $\theta_2$ are the two
angles which describe the $O(4)$ rotation of vectors by parallel
transport around the loops (see \cite{vacuum,wilson}).

As we pointed out in Section 3, the B-G formula is sensitive to such
pure gauge modes, and gives in fact to perturbative order $\hbar G$
the right result for the potential. On the other hand, the T-G
formula would clearly give no potential energy in flat space: if there
is no curvature, the difference between the lengths of the trajectories
of the two masses vanishes. So the two equations ``work'' in a very
different way.

In lattice theory, a version of the T-G formula is employed. In their
non-perturbative simulations, Hamber and Williams \cite{hw} trace two
parallel lines with reference to a (dynamical) Regge lattice and then
compare the lengths $\sum_i l_i$ of the two lines. As we mentioned
in the Introduction, the first results are positive.

In general, the lattice formulation seems to be much more appropriate
for handling quantum gravity than the perturbative expansion. It also has
the property of being automatically coordinate invariant, and of
generating dynamically the flat space limit. The perturbation theory
on flat background makes sense instead as an effective theory for small
$G$, where $G$ is the effective value of the gravitational constant as
measured from the macroscopic Newton force.

In the phase space of lattice theory, specified by the parameters
$a$ (the coupling to the $R^2$-term) and $G_{\rm Bare}$, there is a
continuous line which separates a physical ``smooth'' phase and
a collapsed ``rough'' phase \cite{hw}. The true continuum theory is the
limit of lattice theory on a point of this line, chosen in such a way
that the effective long-distance Newton constant, computed through the
lattice version of T-G formula, equals the measured value.

Then we may argue that a ``correspondence principle'' between the
B-G and T-G formulas requires that $G_{\rm Bare}$ is not as small as
the effective value. Namely, for the region of very small $G_{\rm Bare}$
first order perturbative computations are reliable, so we know there is
no curvature in that regime, and consequently the lattice evaluation
of T-G formula will give $G=0$ exactly (or better, apart from terms
of order $G_{\rm Bare}^2$). On the contrary, if
$G_{\rm Bare}$ is not so small, its effective
value can emerge from lattice theory with no contradiction with the
background computations. Whether this is really what happens, it should
become clear as the simulations proceed.

\m \m \m

It is a pleasure for me to thank D.\ Maison for the warm hospitality in
Munich, and the A.\ Von Humboldt Foundation for financial support.
Also part of this work was made at M.I.T., Center for Theoretical Physics,
and the help of R.\ Jackiw is gratefully acknowledged.

\end{document}